\documentclass[10pt,twocolumn]{article}

\usepackage[a4paper,margin=1.8cm]{geometry}
\usepackage{newtxtext,newtxmath}
\usepackage{lipsum}
\usepackage{xcolor}
\usepackage{mathtools}
\usepackage{multirow}
\usepackage{graphicx}
\usepackage{caption}

\usepackage{hyperref}
\usepackage{cite}

\usepackage{authblk}

\title{High system efficiency nonlinear frequency conversion on thin-film lithium niobate}
\author[1]{Philipp Lohmann}
\author[1,2]{Daniel Wendland}
\author[3]{Francesco Lenzini}
\author[1,*]{Wolfram H.P. Pernice}

\date{}
\affil[1]{Kirchhoff-Institute for Physics, Heidelberg University, Heidelberg, Germany}
\affil[2]{Institute of Physics and Center for Nanotechnology, University of M\"unster, 48149 M\"unster, Germany}
\affil[3]{Consiglio Nazionale delle Ricerche - Istituto di Fotonica e Nanotecnologie (CNR-IFN), Milan, Italy}

\affil[*]{wolfram.pernice@kip.uni-heidelberg.de}
\begin{document}

\twocolumn[
  \begin{@twocolumnfalse}
    \maketitle
    \begin{abstract}
Integrated photonic platforms can greatly enhance the efficiency of nonlinear frequency conversion processes by tightly confining light on a sub-micron scale. However, this advantage is often reduced by large fiber-to-chip coupling losses which drastically reduce the overall performance. Here we demonstrate a highly efficient thin-film lithium niobate frequency converter based on periodically poled waveguides combined with direct laser written out-of-plane couplers. Including on-chip and fiber-to-chip losses we obtain a conversion efficiency of 152 \%/W, thus demonstrating a promising approach for future scalable integrated devices. 
    \end{abstract}
    \vspace{1em}
  \end{@twocolumnfalse}
]

\section{Introduction}

Frequency converters are widely used in both classical and quantum photonics, including, e.g., the realization of coherent sources with broad spectral coverage \cite{doi:10.1126/science.abj4396} or the generation of quantum states of light with non-linear optical processes \cite{PhysRevLett.109.147404,Bouwmeester1997}. Many of these applications rely on the use of materials with a $\chi^{(2)}$ nonlinearity, which can efficiently enable second-harmonic, sum and difference frequency generation processes (SHG, SFG, and DFG) \cite{10.5555/1817101}. 
\noindent With a second-order nonlinear optical coefficient of $\mathrm{d}_{33} \simeq 25$ $\mathrm{pm/V}$ and a transparancy window spanning from UV to mid-infrared wavelengths, lithium niobate (LN) is one of the most commonly used and suitable $\chi^{(2)}$ materials \cite{Zhu:21,Weis1985}. 
The majority of conventional commercial frequency converters rely on quasi-phase-matching (QPM) based on periodic poling in lithium niobate waveguides created via proton exchange or Ti-indiffusion \cite{10.1063/1.326568,Korkishko:98}. Robust fabrication, large and controllable poling periods and facilitated coupling due to large waveguide cross-sections  lead to high conversion efficiencies > 90 \% whereas the normalized conversion efficiencies are < 170 $\%/\mathrm{W cm^2}$  \cite{5446448,Berry:19,Roussev:04,Suntsov:21,Parameswaran:02,Parameswaran:02_2}.\\
\noindent In such waveguides, several Watts of input power and devices with a length of several centimeter are required to compensate for the low normalized efficiency of the intensity-driven nonlinear process, which results from the low refractive index contrast of $\approx 0.02$ and thus a weak modal confinement
\cite{CAI2017405}. This strongly limits the scalability and integrability of those devices. \\

\noindent With the development of thin-film lithium niobate (TFLN) a new platform has emerged on which light can be confined on a sub-micrometer level. This results in compact circuitry, and, together with QPM techniques, in highly efficient nonlinear frequency conversion processes. Although challenging due to very small periods on the micrometer scale, many methods for periodic poling like electric field poling (EFP), focused ion beam \cite{krasnokutska2021submicrondomainengineeringperiodically,Chezganov25042020} , electron-beam-lithography (EBL) \cite{10.1063/1.1575918} or laser writing \cite{10.1063/5.0235673} have been developed for TFLN. EFP as the most common technique has paved the way for various applications including nonlinearities for classical computing \cite{LiSekineNehraGrayLedezmaGuoMarandi+2023+847+855}, all optical switches \cite{Guo2022}, high extinction electro-optical modulators \cite{Sabatti:24} and photon pair generation or squeezed light sources \cite{Shi2024,PhysRevLett.124.163603,chapman2024onchipquantuminterferenceindependent,doi:10.1126/science.abo6213,doi:10.1126/sciadv.adl1814,Xin:22,arge2024demonstrationsqueezedlightsource}. 

\noindent The nonlinear conversion efficiency is commonly benchmarked by SHG measurements, where experiments with single-pass, straight periodically poled TFLN waveguides have already enabled the achievement of normalized conversion efficiency up to 2500-5000 $\%/\mathrm{W cm^2}$ \cite{Wang:18,Chang:16,Boes:19,Zhao:20,Rao:19}. \\

\noindent Although TFLN allows for extremely efficient nonlinear processes, the effectivity of light conversion faces two main challenges. Intrinsic fluctuations in the TFLN film heights have a large impact on the phase-matching condition of the strongly dispersive waveguide modes, which drastically reduces the performance for larger devices. This challenge was tackled in recent years by either keeping the device length small but increasing the optical power locally with ring resonators \cite{Lu:19,Chen:19,Lu:20} or by locally optimizing the phase-matching condition to counteract the intrinsic TFLN film inhomogeneities \cite{mi15091145,Chen2024,Li:24}. \\
\noindent In addition, sensitive fiber-to-chip alignement and lossy coupling solutions based on butt-coupling or grating coupling drastically reduce the overall performance of TFLN-based frequency converters making them impractical for packaged fiber-coupled solutions. \\
\noindent Highly efficient fiber-to-chip couplers are especially important for applications in quantum photonics, where any source of loss can either drastically reduce the photon coincidence count rate, or, when working with squeezed light \cite{Loudon01061987}, introduce high levels of noise in the quantum states.   \\

\noindent In this work, we address the latter challenge and demonstrate a SHG device which exploits the advantage of periodically poled TFLN waveguides in combination with low loss surface couplers based on direct laser written polymer structures. Measured from fiber to fiber, the system efficiency from the near infrared (NIR) to the visible (VIS) is 152 \%/W, which represents an improvement of approximately two orders of magnitude over comparable works. In contrast to butt coupling, our surface couplers enable flexible design architectures, large-scale prototyping and small footprint circuitry. In combination with the above mentioned optimization methods our approach holds promise for highly efficient frequency converters in the future.  
 
\section{Design}
\begin{figure}[ht]
\centering
\includegraphics[width=\linewidth]{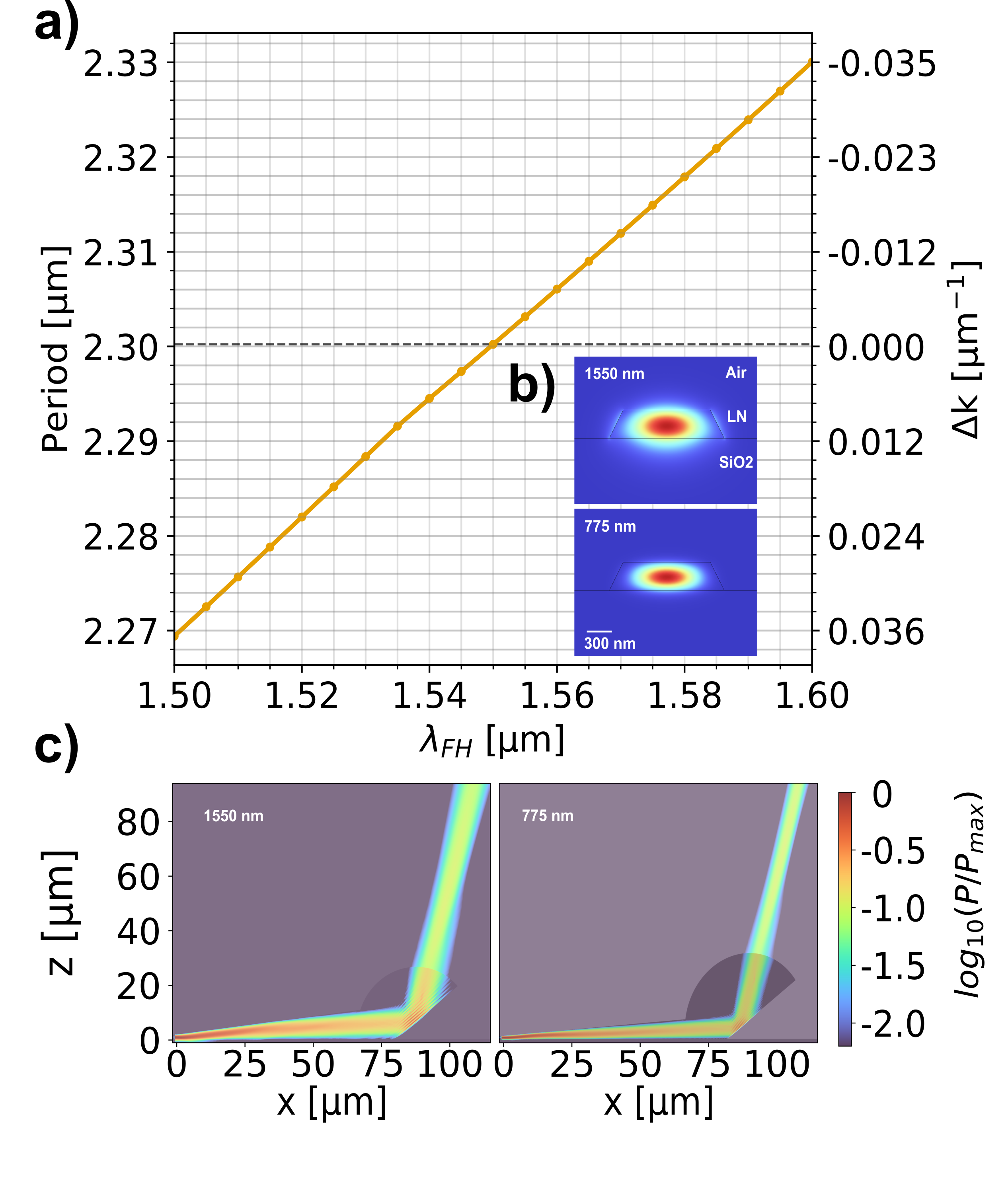}
\caption{ \textbf{a)} The orange curve shows the calculated poling period as a function of the FH wavelength. The period is designed for SHG at 1.55 µm wavelength (grey line) and the resulting momentum mismatch is presented on the right scale. \textbf{b)} Intensity profiles of the fundamental TE modes supported by the waveguide at the FH and SH wavelengths. \textbf{c)} FDTD simulations of the optical powers propagating in the laser-written fiber-to-chip couplers at the FH and SH wavelengths. }
\label{fig:0}
\end{figure}

Our design focuses on the realization of QPM and optimization of the couplers for simultaneous coupling of both wavelengths. QPM is obtained by EFP based on analytical considerations for periodic poling, where the period $\Lambda$ is calculated by 
\begin{equation}
\label{phasematch}
    \Lambda = \frac{\lambda_{\mathrm{FH}}}{2(n_{\mathrm{SH}}-n_{\mathrm{FH}})}
\end{equation}
with the wavelength $\lambda$ and the effective refractive index $n$. Within this work, the subscript $\mathrm{FH}$ and $\mathrm{SH}$ refer to the first and second harmonic, respectively. The geometry relies on a fully etched 300 nm X-cut TFLN waveguide on top of silicon dioxide (SiO2) with a sidewall angle of $\approx 63$°. The waveguide width is chosen equal to approximately 1 micron, as a trade off between a small effective modal area, and a waveguide geometry able to tightly confine the guided mode and achieve low propagation loss due to sidewall scattering. According to Eq. \ref{phasematch}, the poling period was determined using effective indices obtained by the mode solver from COMSOL. In Fig. \ref{fig:0} \textbf{a)} the resulting period is shown as a function of $\lambda_{\mathrm{FH}}$. For a given period, the resulting SHG power spectrum around the resonance is described by 
\begin{equation}
\label{spectrum}
     P_{\mathrm{SH}} (\lambda_{\mathrm{FH}})\propto \mathrm{sinc}^2(\Delta k(\lambda_{\mathrm{FH}}) L /2) 
\end{equation}

where $P$ is the power, $L$ is the poling length and $\Delta k$ is the momentum mismatch, with $\Delta k (\lambda_{\mathrm{FH}})=  \frac{4\pi(n_{\mathrm{SH}}-n_{\mathrm{FH}})}{\lambda_{\mathrm{FH}}}-\frac{2\pi}{\Lambda}$ \cite{10.5555/1817101}. The latter is shown on the right axis in Fig. \ref{fig:0} \textbf{a} for a period of 2.3 µm which is chosen for SHG around 1550 nm. \\
\noindent The corresponding intensity distribution of the waveguide's eigenmodes at the FH and SH wavelengths are visualized in Fig. \ref{fig:0} \textbf{b)}.

\noindent In the non-depletion regime where only a small fraction of the light is converted, the pump can be assumed to have constant power over the poling region resulting in the quadratic dependency
\begin{equation}
    P_{\mathrm{SH}}=\eta_{\mathrm{SHG}}P^2_{\mathrm{FH}}
\end{equation}
where the conversion efficiency $\eta_{\mathrm{SHG}}$ is introduced.
In order to include the coupling and on-chip losses, the formula is adjusted to 

\begin{equation}
\label{deviceefficiency}
    P_{\mathrm{SH,Fiber}}=\eta_{\mathrm{sys}}P^2_{\mathrm{FH,Fiber}}
\end{equation}
with the corresponding powers measured in the fibers. The system efficiency $\eta_{\mathrm{sys}}$ considers all additional losses and can be written as

\begin{equation}
\label{efficiency_relation}
\eta_{\mathrm{sys}} = \eta_\mathrm{SHG}\cdot \eta_{\mathrm{Coupling,FH}}^2\cdot \eta_{\mathrm{Coupling,SH}}
\end{equation}
where the coupling losses $\eta_{\mathrm{Coupling}}$ are explicitly taken into account. All additional losses including propagation losses, mode conversion or bending losses are included in $\eta_\mathrm{SHG}$.

\noindent Using the same waveguide geometry, direct laser written couplers are optimized for both wavelengths. Following our previous work \cite{Gehring:2019_2}, they consist of a lens for beam shaping, a plane for total internal reflection that allows for efficient out-of plane coupling, a taper for beam expansion and a mode converter optimized for the TFLN platform. Corresponding FDTD simulations of the optical power propagating through the couplers are presented in Fig. \ref{fig:0} \textbf{c)} and \textbf{d)}, for NIR and VIS wavelengths, overlayed with the coupler geometry respectively. The coupler geometries were optimized for achieving a high modal overlap with corresponding single-mode fibers, while paying attention to identical focus positions, as well as identical emission angles to allow the usage of a v-groove fiber-array consisting of both fiber types.

\begin{figure}[t]
\centering
\includegraphics[width=\linewidth]{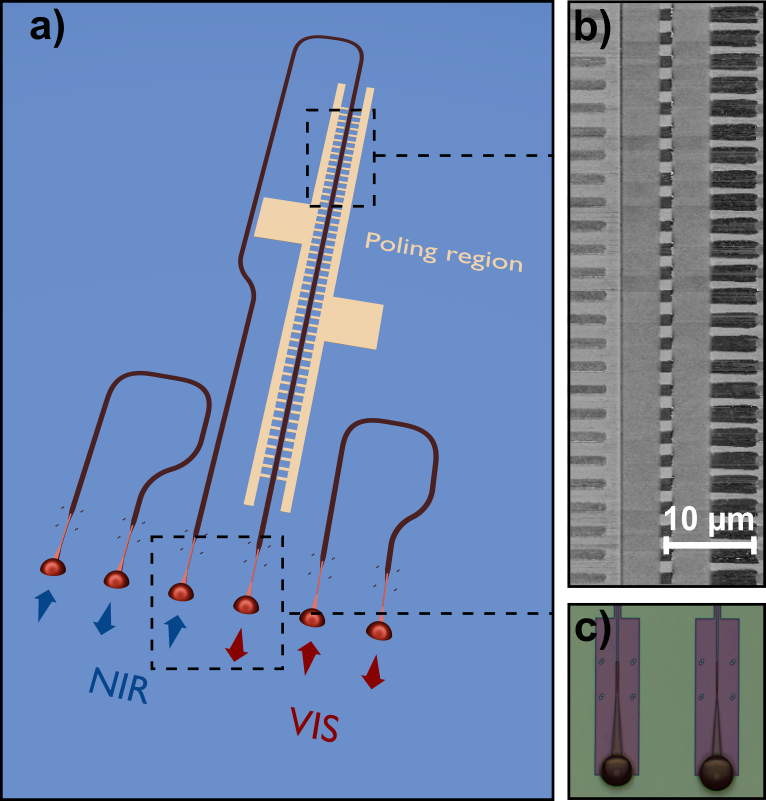}
\caption{\textbf{a)} A scheme of the final device design. A 7 mm poling region is combined with optimized couplers for the FH and SH wavelengths, respectively. Two backloops are used for alignment. \textbf{b)} The resulting periodic poling is investigated using a PFM. Dark grey periodic fingers  on the left and right side represent the gold structures used for EFP. The light grey region around the gold is the remaining LN whereas black regions arising from the right electrodes visualize the poled region. Surrounded by fully etched trenches, the waveguide shows periodic poling close to the desired 50\% duty cycle. \textbf{c)}  A microscope image of the direct laser written surface couplers. The different lens sizes are used for optimized coupling for the different wavelengths and fiber mode field diameters.}
\label{fig:1}
\end{figure}

\noindent The final design of the device is illustrated in Fig. \ref{fig:1} \textbf{a)}. Backloops for both wavelengths are used for facilitating the alignment of the chip to optical fibers. The main device in the center uses two different couplers, one specifically optimized for NIR and the other one for VIS. The poling region is 7mm long which results in a device length of $\approx 1.5$ cm. 
\section{Fabrication}
\noindent Waveguides are patterned on 300 nm thick X-cut LN films bonded on a silicon dioxide substrate, produced from NanoLN.

\noindent As a first step, electrodes for electric field poling are fabricated on the plain LN film. The electrodes are patterned by EBL (100 kV Raith EBPG5150) with a positive resist, followed by physical vapor deposition of a 80 nm thick gold film on top of a 5 nm thick Cr adhesion layer using a lift-off process. Poling of the desired region is performed before fabrication of the photonic circuitry for homogeneous inversion of the ferroelectric domains through the whole depth of the film. With a gap between the electrodes of 15 µm and ten sub-millisecond pulses of 395V homogeneous poling is achieved over the whole length of the device. 
The waveguides are fabricated using positive resist (AR-P 6200.13), EBL and an argon-based etching process in an Oxford Instruments PlasmaPro 100 ICP system. Additional RCA-1 clean and annealing steps are performed for removing the redeposition of sputtered material on the waveguide sidewalls, and reducing the crystal absorption loss, respectively. As the last step, the fiber-to-chip couplers are fabricated by direct laser lithography with a Nanoscribe QX and a commercially available resist (IP-Dip2).  \\
\noindent During the EFP step and after the whole fabrication process, poling is monitored by piezeresponse-force microscopy (PFM), which results in high resolution images of the crystal orientation \cite{10.1063/1.5143266,Jungk_2009,Soergel_2011,cryst11030288}. In Fig. \ref{fig:1} \textbf{b)} a PFM picture of the piezoelectric response's phase for a 60 µm long section of the fabricated waveguide is shown. The poling electrodes are visible on the left and right side on top of TFLN. The LN color indicates the crystal direction and poled domains emerge from the right electrode. The waveguide surrounded by fully etched trenches shows a homogeneous poling region close to the desired 50\% duty cycle. \\
\noindent An optical microscope image of the final coupler region with printed couplers is shown in Fig. \ref{fig:1} \textbf{c)}. The different lens sizes result from individual optimization for the different wavelengths and fiber mode sizes.  
\begin{figure}[h!]
\centering
\includegraphics[width=\linewidth]{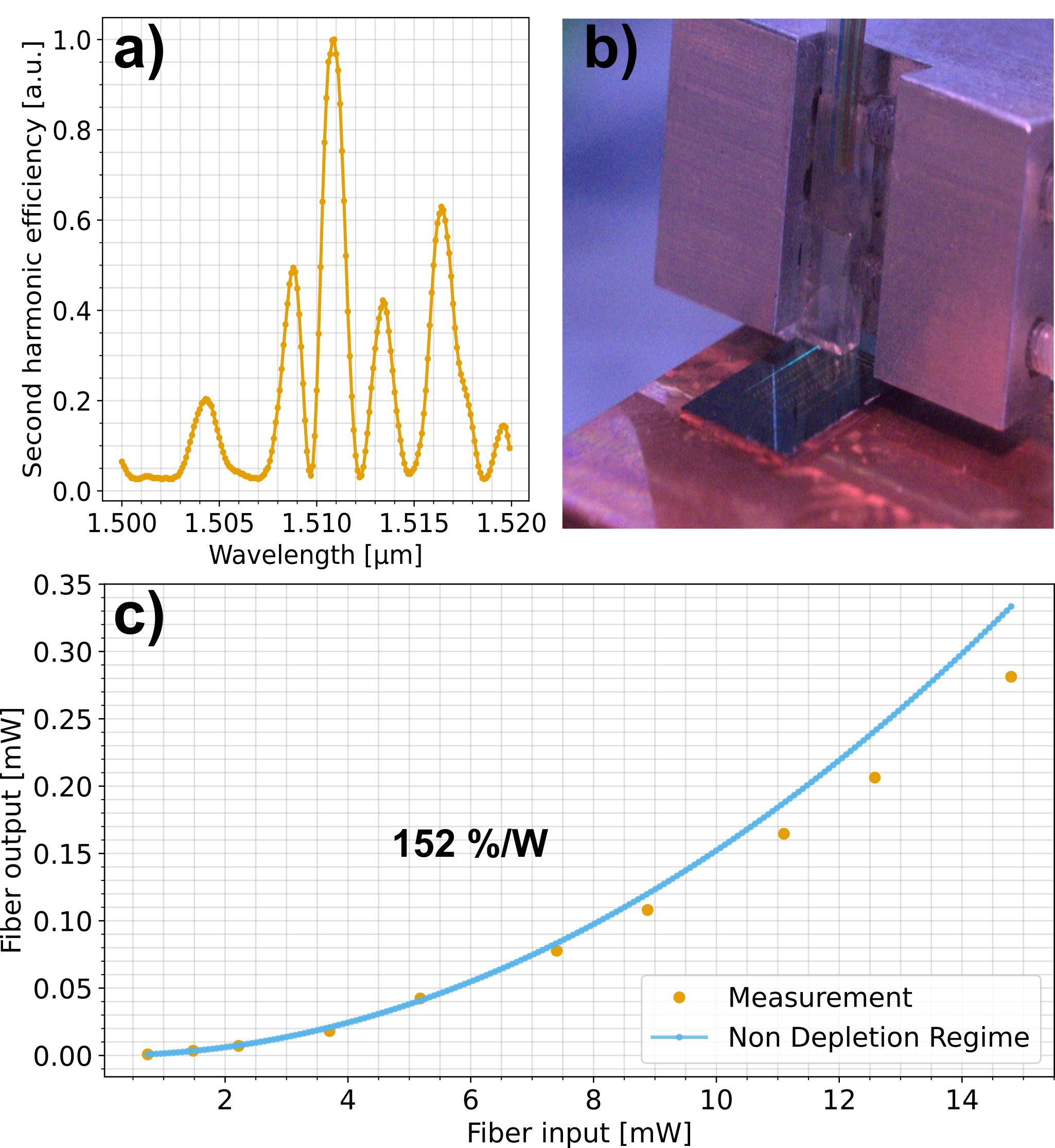}
\caption{\textbf{a)} The resulting second harmonic efficiency is  measured by applying a wavelength sweep. Several peaks are visible with the highest conversion at 1511 nm pump light. \textbf{b)} A camera image of the SHG light. The chip lies on top of a temperature controlled copper holder while light is coupled via a commercial fiber array from the top surface. The SHG light is clearly visible. Whereas by eye, the red light at around 755nm wavelength is directly obtained, the camera perceives it as false-color blue light. \textbf{c)} The SHG power in the fiber is measured as a function of pump power in the fiber. A quadratic fit is applied to the first few data points to obtain the conversion efficiency in the non depletion regime yielding 152 \%/W. }
\label{fig:2}
\end{figure}

\section{Results}
The measurements were performed with a custom-made fiber array unit equipped with both SMF28 and 780HP fibers for the infrared and visible regime respectively. A Santec TSL-770 tunable laser source is used as a pump and a fiber-coupled polarization controller is added to maximize coupling in the TE0 mode. The transmission is measured using photodetectors for both wavelengths. The chip is temperature stabilized to $23$ °C by using a thermo-electric cooler (TEC) and the alignment to the fiber array is optimized by measuring the transmission through the two backloop structures on both sites of the device of interest. Additionally the backloops are used to extract the coupling losses for both wavelength respectively. We measure $\eta_{\mathrm{Coupling,FH}} = 0.74$ corresponding to $\approx -1.3$ dB/coupler and $\eta_{\mathrm{Coupling,SH}} = 0.516 \approx -2.9$ dB/coupler. The backloop's coupler transmission for the SH is measured while optimizing the signal to the FH in order to obtain values most comparable to the frequency converter device. \\
\noindent The wavelength dependent converted power is measured by sweeping the pump wavelength and the resulting second harmonic efficiency is reported in Fig. \ref{fig:2} \textbf{a)}. In comparison to Eq. \ref{spectrum} a deviation from a perfect sinc-like shape can be seen, which can be attributed to inhomogeneities in the LN film and is a common problem for devices of several mm length \cite{Chen2024}. The maximum of the conversion is found around 1511 nm pump wavelength which deviates from the desired wavelength of 1550 nm. In addition to the film thickness, the waveguide width plays a crucial role for the SHG resonance and resist shrinkage next to slight fluctuations in the fabrication process can lead to those deviations. A camera image of the chip during the experiment is shown in Fig. \ref{fig:2} \textbf{b)}. It is mounted on a copper holder which is used in combination with the TEC as a temperature controller. A single fiber array unit positioned in close proximity to the surface of the chip is used for in-coupling of the pump light, as well as for out-coupling of the light created by SHG. The converted light, building up as it travels through the periodically poled region, is clearly visible. \\

\noindent After setting the pump wavelength to $1511$ nm for maximizing the conversion, a power sweep is performed and the measured SHG power as a function of the pump power is reported in Fig. \ref{fig:2} \textbf{c)}. By fitting the data with a quadratic function (see Eq. \ref{deviceefficiency}) a normalized conversion efficiency $\eta_{\mathrm{sys}}=152$ \%/W is estimated.\\ 
\noindent The on-chip conversion efficiency can be calculated by including the coupling losses in Eq. \ref{efficiency_relation}, which yields $\eta_\mathrm{SHG} = 538 \%$ /W. \\
\noindent Finally, we test the thermal behavior of our device and obtain a resonance tunability of 0.1 nm/K by heating the whole chip with the TEC.

\noindent In Table \ref{tab:comparison}, the results obtained in this letter are compared to other recent works, which utilize TFLN for efficent SHG. The list is restricted to works, where values for $\eta_{\mathrm{Coupling,FH}}$ and $\eta_{\mathrm{Coupling,SH}}$ are explicitly stated. The results obtained in this letter substantially outperform similar works based on single-pass waveguides with constant period, where a maximum system efficiency of only 1.1 \%/W is achieved. \\
\noindent Optimizing the phase-matching condition locally by adaptive poling requires far more in-depth optimization and yields on-chip efficiency ten times larger and an overall system efficiency of 162 \%/W. Note that our work achieves a comparable system efficiency with a waveguide which is 3 times shorter, and without the need of relying on complex adaptive optimization procedures. \\
\noindent SHG devices based on double-resonant rings yield exceptionally high system efficiencies up to 9941 \%/W. However, a drawback of resonant structures is that they are affected by a narrow linewidth typically of the order of only a few hundred of MHz, making them unsuitable for several applications e.g., ultra-fast classical and quantum information processing \cite{li2025allopticalcomputing100ghzclock,doi:10.1126/science.abo6213,10.1063/5.0137641} where much broader bandwidths are required. Moreover, in double-resonant ring resonators it is not possible to achieve an independent tuning of the resonance positions of FH and SH modes. As nanoscale deviations in the fabrication lead to different resonant wavelengths for rings patterned on the same chip, it becomes highly challenging to scale-up these devices.\\
\begin{table*}[htbp!]
\centering
\caption{\bf Performance comparison of different TFLN SHG devices.}
\begin{tabular}{|c|c|c|c|c|c|c|c|}
\hline
Type & Reference & Length (mm) & $\eta_{\mathrm{SHG}}$ ($\%/W$) &$\eta_{\mathrm{Coupling,FH}}$ & $\eta_{\mathrm{Coupling,SHG}}$ &$\eta_{\mathrm{sys}}$ ($\%/W$)  \\
\hline \hline
\multirow{6}{9em}{Single waveguide and constant period}&\cite{Chang:16}  & 4.8 & 37 & 0.25 & 0.04 & 0.03\\
&\cite{Wang:18}& 4.0 & 416 & 0.1 & 0.1 & 0.42\\
&\cite{Rao:19}   & 0.6 & 17 & 0.25 & 0.2 & 0.2\\
&\cite{Zhao:20}   & 5 & 939 & 0.17 & 0.04 & 1.1\\
&\cite{Sabatti:24}   & 7 & 35 & 0.4 & 0.2 & 1.1\\
& This work  & 7 & 538 & 0.74 & 0.516 & 152\\ \hline 
Adapted poling &\cite{Chen2024}   & 21 & 9500 & 0.29 & 0.2 & 162\\ 

\multirow{3}{*}{Ring resonator} & \cite{Lu:19}& 0.44 & 250000& 0.126& 0.055&218 \\ 
& \cite{Chen:19}& 0.3 & 230000& 0.282& 0.141&2579 \\ 
& \cite{Lu:20} & 0.44 & 5000000 & 0.14&0.1& 9941 \\\hline 
\end{tabular}
  \label{tab:comparison}
\end{table*}



\section{Conclusion}
Whereas the advantage of TFLN for $\eta_{\mathrm{SHG}}$ was demonstrated in prior works, we focus on the overall system performance, which suffers from local inhomogeneities  and poor fiber-to chip coupling efficiency. By focusing on the coupling optimization, we demonstrate a frequency converter for SHG based on TFLN integrated circuitry yielding a fiber-to-fiber conversion efficiency of 152 \%/W, which is two orders of magnitude higher than similar works and comparable to the approach with adapted poling. \\
\noindent With our direct laser written surface couplers, low-footprint TFLN circuitry and periodic poling, our approach holds great promise for the realization of highly efficient, broadband and scalable frequency converters for both classical and quantum photonic applications. \\
\noindent For future applications our device is completely compatible with local poling optimization techniques like local thermal tuning, adaptation of the poling period or adaption of the waveguide width, which can be used to ensure constructive phase-matching over centimeter of poling lengths, yielding an additional tenfold increase in conversion efficiency.


\section*{Funding}

The authors acknowledge the support of the European Union’s Horizon Europe research and innovation program under Grant Agreement No. 101135288 (EPIQUE) and No. 101080173 (CLUSTEC)

\smallskip

\section*{Disclosures} The authors declare no conflicts of interest.

\section*{Data availability} Data underlying the results presented in this paper are not publicly available at this time but may be obtained from the authors upon reasonable request.

\bibliographystyle{unsrt}
\bibliography{refs}

\end{document}